\documentclass[aip,jap,reprint,twocolumn,floatfix]{revtex4-1}
\usepackage{dcolumn}
\usepackage{graphicx}
\usepackage{epsfig}
\usepackage{bm}
\usepackage{amsmath}
\usepackage{amssymb}
\usepackage{lmodern}

\begin{document}

\title{A two-in-one superhydrophobic and anti-reflective nanodevice in the grey cicada \emph{Cicada orni} (Hemiptera)}

\author{Louis Dellieu}
\email{louis.dellieu@unamur.be}
\affiliation{$^{1}$Research Center in Physics of Matter and Radiation (PMR), Department of Physics, University of Namur (FUNDP), 61 rue de Bruxelles, B-5000 Namur, 
Belgium}

\author{Micha\"{e}l Sarrazin}
\email{michael.sarrazin@unamur.be}
\affiliation{$^{1}$Research Center in Physics of Matter and Radiation (PMR), Department of Physics, University of Namur (FUNDP), 61 rue de Bruxelles, B-5000 Namur, 
Belgium}

\author{Priscilla Simonis}
\affiliation{$^{1}$Research Center in Physics of Matter and Radiation (PMR), Department of Physics, University of Namur (FUNDP), 61 rue de Bruxelles, B-5000 Namur, 
Belgium}

\author{Olivier Deparis}
\affiliation{$^{1}$Research Center in Physics of Matter and Radiation (PMR), Department of Physics, University of Namur (FUNDP), 61 rue de Bruxelles, B-5000 Namur, 
Belgium}

\author{Jean Pol Vigneron}
\altaffiliation{Our friend and colleague Prof. Jean Pol Vigneron accidentally died on June 24, 2013, during the early writing of this article.}
\affiliation{$^{1}$Research Center in Physics of Matter and Radiation (PMR), Department of Physics, University of Namur (FUNDP), 61 rue de Bruxelles, B-5000 Namur, 
Belgium}

\begin{abstract}
Two separated levels of functionality are identified in the nanostructure which covers the wings of the
grey cicada \textit{Cicada orni} (Hemiptera). The upper level is responsible for
superhydrophobic character of the wing while the lower level enhances its
anti-reflective behavior. Extensive wetting experiments with various chemical species and optical measurements were performed in order
to assess the bi-functionality. Scanning electron microscopy imaging was
used to identify the nanostructure morphology. Numerical optical simulations
and analytical wetting models were used to prove the roles of both levels
of the nanostructure. In addition, the complex refractive index of the chitinous material of the wing was determined from measurements.
\end{abstract}


\maketitle

\section{Introduction}

For a long time, anti-reflective properties have a major importance in the
development of optical devices. For instance, anti-reflective coatings are
required when designing telescopes, camera lenses, or glass windows. Such
coatings allow e.g. avoiding undesirable reflections in optical imaging \cite
{37}. Moreover, since anti-reflective coatings are wavelength-dependent,
they can be used to reject efficiently unwanted radiation, for instance to
enhance ultraviolet light protection in sunglasses \cite{37}. In another
context, superhydrophobicity is a key property for numerous industrial
applications \cite{38}. Indeed, superhydrophobic surfaces can exhibit
self-cleaning properties: since water droplets cannot stick on such a
surface, the rolling drops clear the surface from any impurities \cite{26}.
As a matter of fact, an optical device that could combine both
anti-reflective and superhydrophobic properties is highly interesting \cite
{39}.

Anti-reflection is well known in biological organisms: e.g. nanostructures
such as nipple arrays act as an adaptative refractive index layer in moth
eyes for instance \cite{1,2}. On the other hand, superhydrophobicity is also
well-known in many plants \cite{3,4,5,6,7,8,9,10}, for instance in lotus 
\cite{7,8,11,12}. Hydrophobic properties and antireflecting nanostructures
of cicada wings have been studied independently in previous works \cite
{22,23,24,20a,20b,20c}. Although the function as well as the physics of
nipple arrays is well documented from the point of view of either
antireflection \cite{23} or hydrophobicity \cite{24,20a,20b,20c}, the
understanding of the role of the geometry in the interplay between optical
and wetting properties deserves further investigations. Indeed, the
combined aspect of wetting/hydrophobic/anti-reflective properties has not
yet been considered as such, strictly speaking. Hereafter, thanks to
extensive modeling and measurements, we show that the grey cicada combines
both hydrophobic and anti-reflective properties in an entangled two-level
nanostructure. Especially, wetting and hydrophobic properties are studied
for a wide range of liquids with various surface tensions. In addition, we
determine the complex refractive index of the wing, which is useful and
often needed as reference in biophotonics simulations.

\begin{figure}[b]
\centerline{\ \includegraphics[width=8 cm]{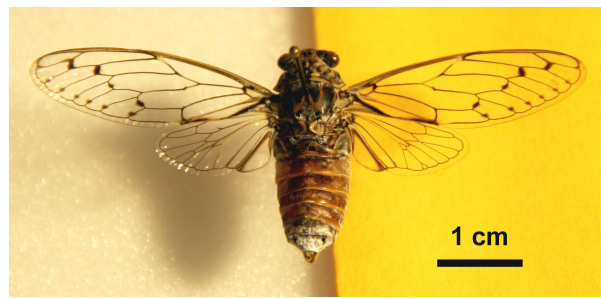}}
\caption{(Color online). The grey cicada \textit{Cicada orni} (Hemiptera).}
\label{fig1}
\end{figure}

\begin{figure}[t]
\centerline{\ \includegraphics[width=8 cm]{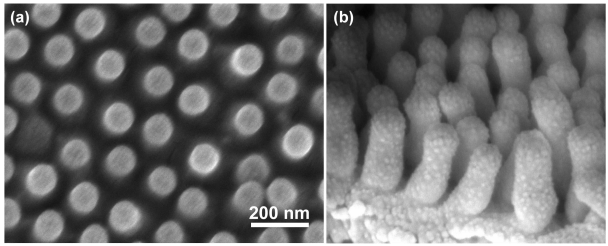}}
\caption{Scanning electron microscopy images of the cicada wing (same scale for both figures). a: Typical
hexagonal nipple array is clearly observed. b: Protuberances consist of
conic-like tips with a hemispherical top.}
\label{fig2}
\end{figure}

\section{Morphology and structural model of the wing membrane}

The grey cicada (Fig. 1) is one of the most frequent cicada species in the
south of France. The studied specimen was captured at Al\'{e}s in the
C\'{e}vennes region. The grey cicada is about $40$ mm long and has a $75$ mm
wing span. In order to investigate the nanostructure of the wing, scanning
electron microscopy (SEM) imaging was performed with \textit{Jeol} $7500$ 
\textit{F} scanning electron microscope.

Both sides of the wing are covered by a hexagonal nipple array (Fig. 2(a)).
Each protuberance is a truncated cone, with a hemispherical top (Fig. 2(b)).
Protuberances lay on a $2.7$ $\mu $m-thick slab. This type of protuberance
is often encountered in compound eyes or corrugated wings of insects \cite
{1,13,18,25}. According to SEM pictures, the typical dimensions and geometry
of the different parts of the structure can be estimated on average (see
Fig. 3(a)). The truncated cone has a circular base of radius $R\approx85$
nm, and a height $h\approx160$ nm. The top hemisphere has a radius $%
r\approx40$ nm. The total height (cone $+$ hemisphere) is therefore $%
H\approx200$ nm. The average hexagonal lattice parameter is $a_0\approx173$
nm. We fairly assume that the disorder in the biological structure, i.e.
weak deviations with respect to perfect lattice, does not significantly
affect the main conclusions of the present study \cite{32}.

\section{Optical properties}

In the following, we highlight the anti-reflective properties of the cicada
wing through scattering and transmission measurements. Based on models of
the wing structure, simulations were performed in order to study the role of
the morphology on the optical properties.

\subsection{Experimental results}

\begin{figure}[b]
\centerline{\ \includegraphics[width=8.5 cm]{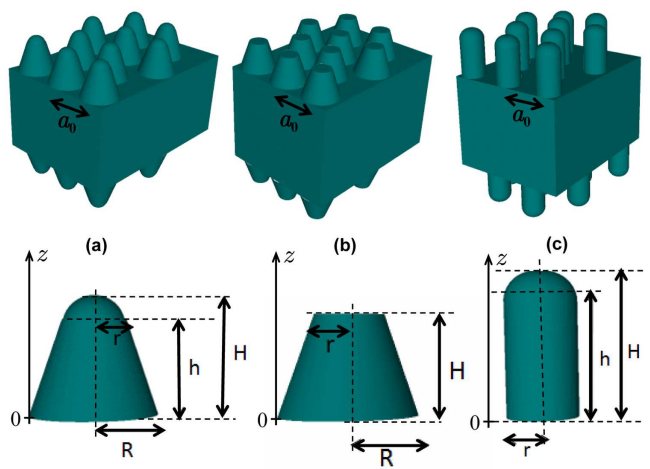}}
\caption{(Color online). Various geometrical models studied. a: Model of the
wing structure, b: Model used to study the influence of top hemisphere on
transmission, c: Model used to study the influence of truncated cone on
transmission.}
\label{fig3}
\end{figure}

The diffusive and polarizing characters of the wing were investigated at
incidence angles \textit{i} of $0\ensuremath{^\circ}$, $15\ensuremath{^\circ}$
and $30\ensuremath{^\circ}$ using scatterometry in transmittance mode (%
\textit{ELDIM EZcontrast} scatterometer). This scatterometer uses Fourier
lenses to record the bidirectional transmittance or reflectance distribution
function (BTDF or BRDF) of the sample (Fig. 4(a)) at visible wavelengths.
In order to detect possible polarization of the emerging light, a polarizer
was placed as analyzer as shown in Fig. 4(a). In a first step (calibration),
at each incidence angle, the BTDF of the incident beam alone (without wing
and polariser) was measured, showing purely specular transmittance as
expected (Fig. 4(b)). In a second step, the wing was inserted and the
polarizer was positioned perpendicular ($\alpha $ $=90\ensuremath{^\circ}$)
to the wing longitudinal axis and then parallel to it ($\alpha =0%
\ensuremath{^\circ}$). BTDFs of the wing for both analyzed polarizations
were recorded. At each incidence angle \textit{i}, both BTDF maps were compared
to the BTDF map of the incident beam alone. As 
polarization effects should be observed preferably at large incidence angles, the
results in Fig. 4(c) and 4(d) are presented for the incidence angle \textit{i} of 
$30$ degrees (and a wavelength $\lambda $ $\cong $ 510 nm in the middle of
visible range). Accordingly, neither diffusive effects nor polarizing
effects of the wing could be detected (Fig. 4(c) and 4(d)). Results (not
shown) were identical for the other wavelengths in the visible spectrum
(according to the characteristics of our scatterometer). Since the wing
exhibited a very low diffusive character and in order to optimize the signal
to noise ratio, transmission properties are mainly emphasized hereafter by
contrast with reflection properties.

Measurement of the wing optical properties spectrum were performed at normal
incidence using a double beam spectrophotometer (\textit{Perkin Elmer Lambda}
$750$ \textit{S} \textit{UV/VIS/NIR}) and an integrating sphere. In Fig.
5(a), reflection spectrum is shown. Since reflection is very weak,
measurement was made with the wing on a black substrate (Certified
Reflectance Standard SRS-02-020 \textit{Labsphere} with a reflectance factor
of 2$\%$ between $250$ and $2500$ nm). The measured reflection $R_m$ had to
corrected since $R_m$ was fairly approximated by $R_m \approx R_w +
R_{bg}\cdot T_w^2$ (see insert in Fig. 5(a)) where $R_w$ was the expected
wing reflection, $R_{bg}$ was the black substrate reflection (which was
measured independently, not shown), and $T_w$ was the free-standing wing
transmission which is measured independently (see hereafter). In Fig. 5(a),
the blue dashed curve is $R_m$ whereas the black solid curve is $R_w$ which
is deduced from the above expression. This correction confirms that
reflection values are weak in accordance with previous results \cite{23},
and must be derived carefully. In addition, it is well-known that
spectroscopic measurements are impacted by instrumental noise for low
transmittance or reflectance levels. As a consequence, it is more relevant
to compare high-level transmittance measurements with numerical models,
rather than low-level reflectance measurements.

Let us now consider transmission measurements. They were carried out on
free-standing wings i.e without substrate. The black curve in Fig. 5(b)
indicates an almost perfect transmission in the visible range although it
appears to be weaker on the blue side than on the red side of the spectrum.
This means that almost 100\% of the light passes through the free-standing
wing without being absorbed or reflected. The transmission exhibits a large
drop below 350 nm, due to a strong intrinsic absorption in the ultra-violet
range. Note that in Fig. 5(a) we found that reflection does not fade away at
these wavelengths. Using the absorption coefficient calculated from
reflection and transmission at each wavelength, the spectrum of the
imaginary part of the complex refractive index was deduced from the
Beer-Lambert law \cite{20} as shown in Fig. 5(c).

In order to reveal the effect of the nipple array on the optical properties,
the wing was covered with a calibrated refractive-index oil. Indeed,
provided that the oil refractive index matches the refractive index of the
protuberances, the oil filling the air cavities and the material can be
regarded as a single planar layer. Here, a decrease of transmission was
noted using a calibrated commercial oil with a real part of refractive index
equal to $1.56$ (see blue dashed curve on Fig. 5(c)). This oil is certified by the manufacturer \textit{Cargille
Labs} (also verified by us) with a negligible dispersion on the visible
range and a weak dispersion in the ultraviolet domain. Such a
refractive-index value is consistent with that of chitin \cite{33}. As a
result, in the following, we will assume that the cicada wing is made of a
chitin-like compound with an average refractive index of $1.56$. On Fig.
5(b), the blue dashed curve (transmission spectrum of wing immersed in index
matching oil) shows a transmission drop of $6$\% in the visible range and
spectral changes in the UV range. One can then conclude that the nipple
array increases the transmission by $6$\%, improving the transparency of the
wing. These results are in agreement with those obtained by other authors 
\cite{23}. Since UV absorption of the chitin material was determined, the
absorption coefficient of the wing (extrapolated from experimental
measurement) (Fig. 5(c)) will be used in the following numerical simulations.

\begin{figure}[!ht]
\centerline{\ \includegraphics[width=8.5 cm]{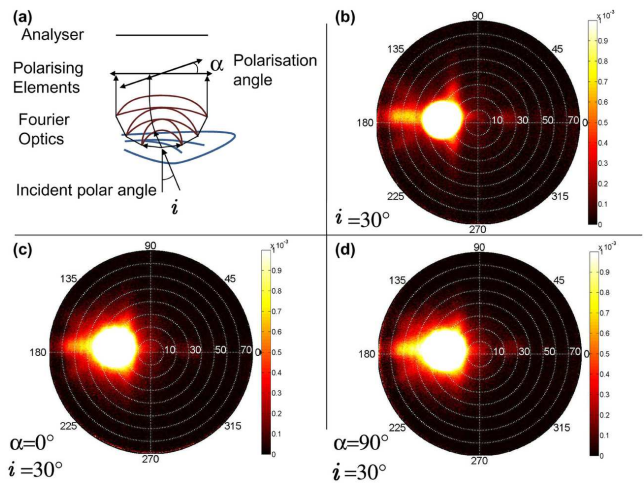}}
\caption{(Color online). Bidirectional Transmittance Distribution Functions
(BTDF) (in watt/m$^2$/sr/nm). Results are shown for polar angles ranging
from $0\ensuremath{^\circ}$ to $80\ensuremath{^\circ}$ and azimuthal angles
in ranging from $0\ensuremath{^\circ}$ to $360\ensuremath{^\circ}$. The
wavelength is $\lambda$ = $509.48$ nm. a: Principle of the scatterometer. b: BTDF of the incident beam at an
angle of incidence $i=30\ensuremath{^\circ}$. c,d: BTDF of the cicada wing
at $i=30\ensuremath{^\circ}$ and for two orthogonal analyzing polarizer
orientations, $0\ensuremath{^\circ}$ (panel (c)) and $90\ensuremath{^\circ}$
(panel (d)).}
\label{fig4}
\end{figure}

\begin{figure}[!ht]
\centerline{\ \includegraphics[width=7.5 cm]{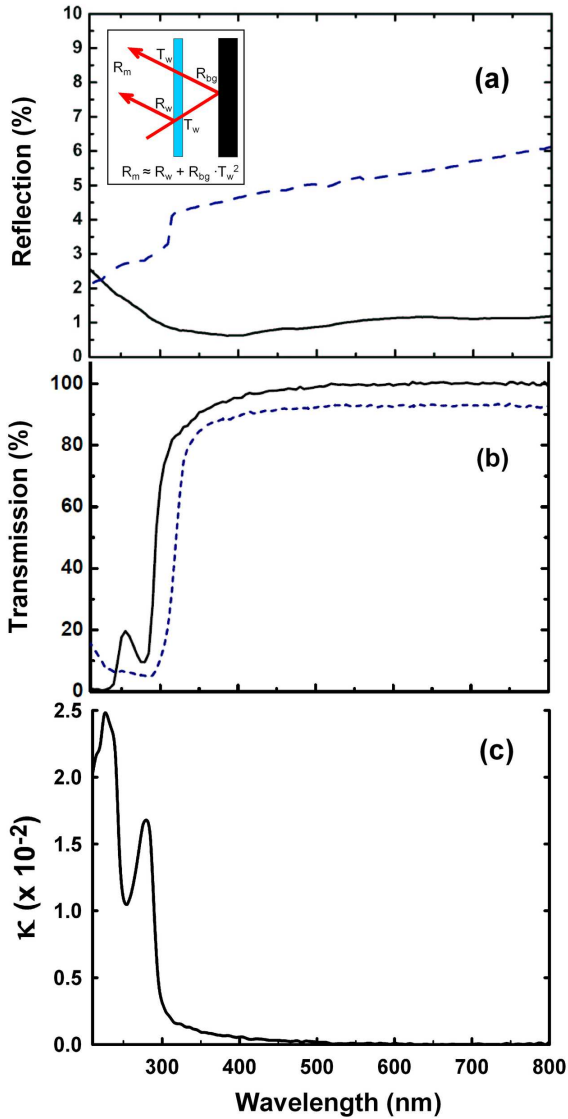}}
\caption{(Color online). a: Measured wing reflection spectrum at normal
incidence (blue dashed curve) and corrected measurement (black solid curve).
Insert: Correction procedure. b: Measured wing transmission spectrum.
Measurements on the actual free-standing wing (black curve) and the wing
immersed in index matching oil (blue dashed curve). c: Imaginary part of the
optical index of the cicada wing obtained from reflection/transmission
measurements.}
\label{fig5}
\end{figure}

\begin{figure}[ht!]
\begin{center}
\includegraphics[width=7.5 cm]{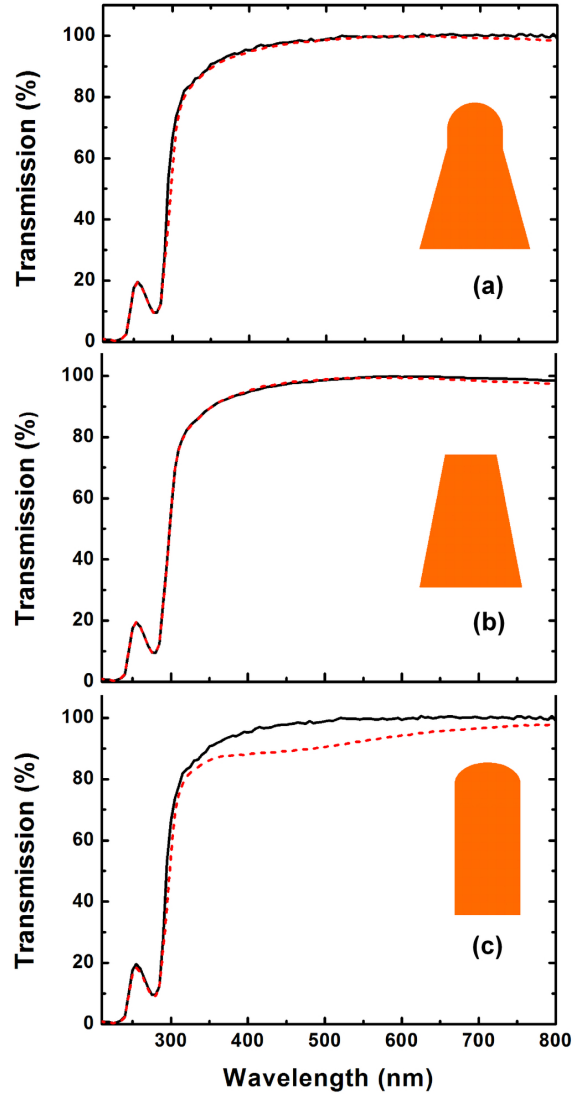}
\end{center}
\caption{(Color online). a: Measurement (black solid curve) and theoretical
simulation (red dashed curve) on the actual wing model. b: Measurement
(black solid curve) vs. theoretical simulation on the wing model without
hemispheres on top (red dashed curve). c: Measurement (black solid curve)
vs. theoretical simulation on the wing model with cylinders instead of
truncated cones (red dashed curve).}
\label{fig6}
\end{figure}

\begin{figure}[ht!]
\begin{center}
\includegraphics[width=8.5 cm]{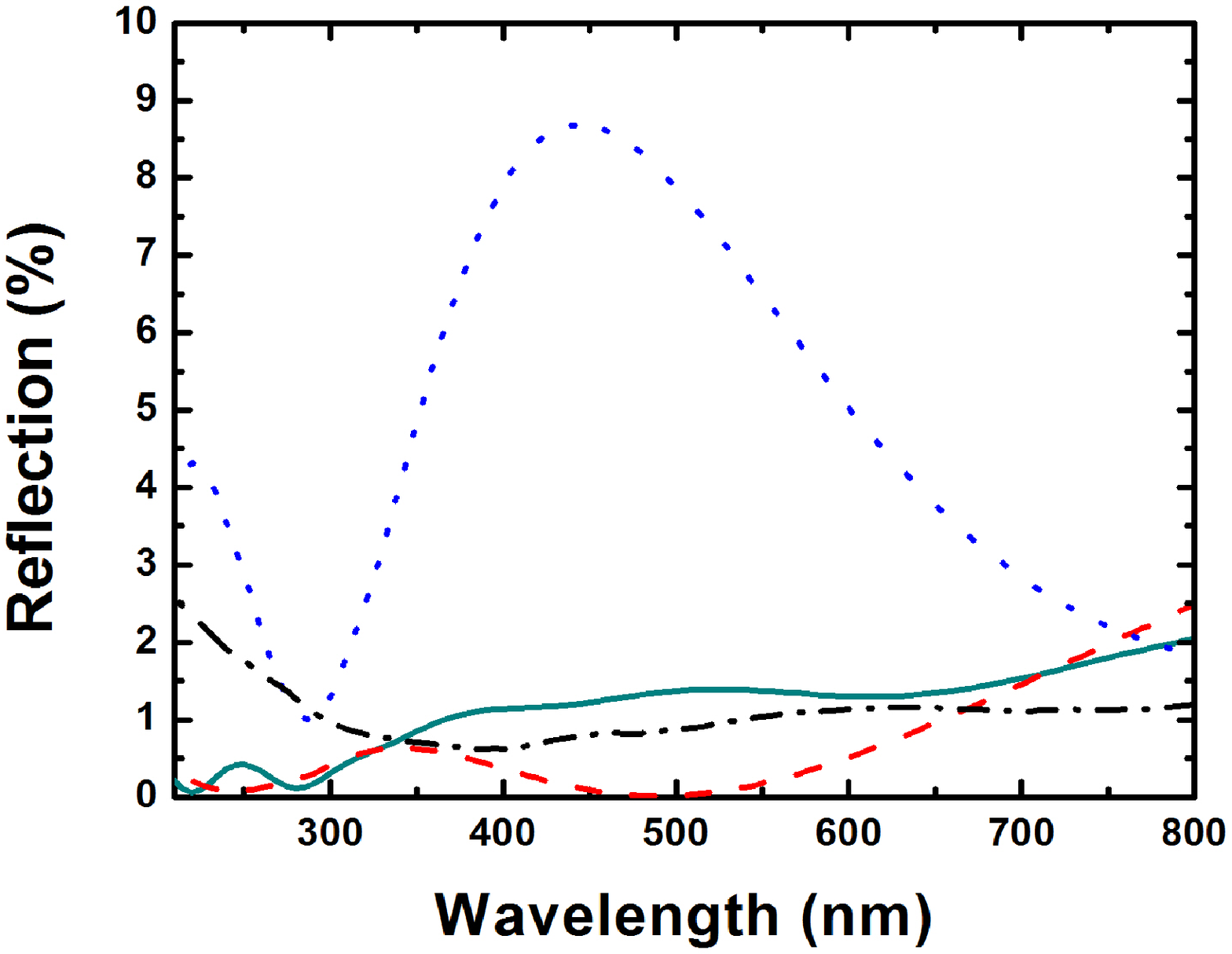}
\end{center}
\caption{(Color online). Theoretical simulation (red dashed curve) of the
reflection on the full wing model (Fig. 3 (a)). This curve can be compared
with the simulated reflections for the wing model without hemispheres on top
(cyan solid curve) or with cylinders instead of truncated cones (blue
short-dashed curve). The black dash-dotted curve is the measured reflection.}
\label{fig7}
\end{figure}

\subsection{Theoretical analysis}

On the basis of morphological data, we described the wing as a homogeneous
slab with, on both side, a hexagonal array of truncated cones with top
hemisphere (Fig. 3(a)). For the sake of simplicity, we assumed that the wing
was made of chitin only. The real part of the refractive index was taken
equal to $1.56$ and was assumed to be constant all over the simulation
spectral range whereas the absorption coefficient was taken
wavelength-dependent according to data shown in Fig. 5(c). Simulations of
the optical properties were carried out from the near UV to the near
infrared ($200$ to $800$ nm) by using a transfer matrix code \cite{32}.

Since the nipple array has subwavelength dimensions, the corrugated layer
can be considered as a continuous effective material with a graded-index
along its thickness. More specifically, the effective dielectric constant
can then be defined by \cite{1}:

\begin{equation}
\varepsilon(z) =\varepsilon _{air}+(\varepsilon _{chitin}-\varepsilon
_{air})f(z),  \label{eff}
\end{equation}
where $\varepsilon _{air}$ is the dielectric constant of air ($\varepsilon
_{air}=1$), $\varepsilon _{chitin}$ is the dielectric constant of chitin and 
$f(z)$ is filling factor given by:

\begin{equation}
\ f(z)=\frac{\pi r^2(z)}S,  \label{ratio}
\end{equation}
with $S=\frac{a_0^2\sqrt{3}}2$ and $r(z)$ the radius of the circular section
of a nipple at coordinate $z$. While light propagation in corrugated layers
is computed thanks to the transfer matrix code, light propagation in the
homogeneous thick slab is calculated using an analytical expression based on
Fresnel equations in which phases are set to zero in order to mimic optical
incoherence. Indeed, structural imperfections in the slab, which is a thick
biological layer, tend to break constructive light wave interferences.
Finally, although we choose to focus on transmission for the reasons we
exposed previously, simulations of reflection are also performed to
corroborate our results.

Simulation results for transmission are shown in Fig. 6(a). The black curve
corresponds to the measured transmission whereas the red curve corresponds
to the simulated transmission: The two curves match very well. This result
is valid whatever the polarization state is. In order to discriminate the
role of the truncated cone and the role of the top hemisphere, two
alternative wing models were considered. We first considered a truncated
cone without the hemisphere on top (see Fig. 3(b)). Fig. 6(b) shows the
simulated spectrum (red curve) and the measured spectrum (black curve). Both
transmission spectra match quite well, showing that the hemisphere does not
actually contribute to the transparency of the wing. In a second simulation,
we considered a cylinder (instead of a cone) (see Fig. 3 (c)) with a
hemisphere on top. The result is shown in Fig. 6(c). The simulated
transmission (red curve) is significantly lower than the measured one (black
curve). Together with the previous simulation, this result shows that the
truncated cone is required to get transparency in the visible range. This
can be easily understood since the truncated cone avoids a sudden change in
the refractive index when light impinges on the slab. Indeed, considering
Eq. (1), the truncated cone creates a gradient of refractive index which
acts as an optical impedance adaptation layer.

Let us now consider the simulated reflection. Our results are summarized in
Fig. 7. The measured (black dash-dotted curve) and simulated (red dashed
curve) reflections exhibit comparable values about $1$\%. Nevertheless, we
note that spectra of both reflections do not perfectly match. As explained
previously, this can be explained by the lack of accuracy when measuring
weak reflections. We also present the simulated reflection for the wing
model without hemispheres (cyan solid curve). As for the transmission (Fig.
6(b)), there is negligible difference (less than $2$\%) with the full wing
model (Fig. 3(a)). Now, by contrast, if we consider the model with cylinders
instead of truncated cones, we see that the reflection increases by a factor 
$10$. This corroborates qualitatively our results obtained with the
transmission (Fig. 6(c)). In addition, we note the strong oscillations of
the simulated reflection, since the array of cylinders acts as a thin
effective homogeneous slab where interferences can occur.

\section{Wetting properties}

In this section, contact angle measurements are reported and wettability
regimes are investigated. Analytical models are used to describe the
superhydrophobicity of the wing.

\begin{figure}[t]
\centerline{\ \includegraphics[width=7.5 cm]{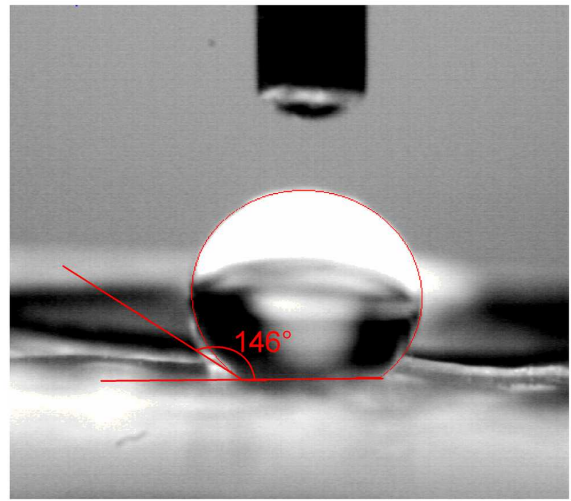}}
\caption{(Color online). Wetting measurement on wing of \textit{Cicada} 
\textit{orni}, the contact angle is $146^\circ$.}
\label{fig8}
\end{figure}

\begin{figure}[t]
\centerline{\ \includegraphics[width=7.5 cm]{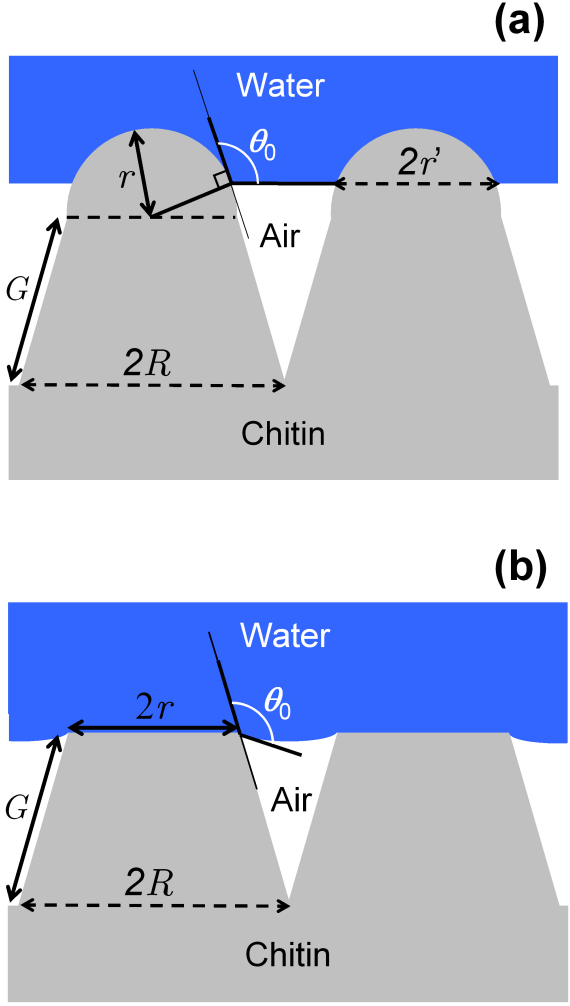}}
\caption{(Color online). Wetting models under consideration. a: Wetting model 
from the full cicada's wing description. b: Wetting model without hemispheric top. The angle $\theta_{0}$ corresponds to the Young's contact angle for a flat chitin surface. }
\label{fig9}
\end{figure}

\subsection{Experimental results}

A surface can be classified as superhydrophobic if the contact angle of a
water droplet on the surface is higher than $150$$^{\circ }$. If the contact
angle lies between 90$^{\circ }$ and $150$$^{\circ }$, the surface is only
hydrophobic and it is hydrophilic if the contact angle lies between $0$$%
^{\circ }$ and $90$$^{\circ }$. In the present case, the contact angle of a
water droplet on the wing surface is measured to be $146$$^{\circ }$ (Fig.
8). As a consequence, the wing of \textit{Cicada orni} is almost
superhydrophobic. According to previous works related to the link between
nanostructure and superhydrophobicity \cite{13,15,16,17,20a}, we can fairly
assess that the nipple array covering the wing surface is responsible for
its hydrophobic properties (see Fig. 9). Indeed, without the nanostructure, the water
droplet would probably spread on the surface since the contact angle of
water on a flat chitin surface is about $105^{\circ }$ \cite{21}.

For a better understanding of the wing hydrophobicity, it is relevant to
measure the wetting properties for various liquids with different surface
tensions. For this purpose, we used ethanol-water solutions with different
mass percentages of ethanol (from 0\% to 100\%), but also a $6.0$ M sodium
chloride aqueous solution and mercury.

The effective contact angle $\theta _{eff}$ for each liquid on the wing (the
nanostructured surface) was measured, and $\cos \theta _{eff}$ was plotted
against $\cos \theta _0 $, where $\theta $ is the contact angle on a flat
chitin surface (Fig. 10 ; black dots correspond to measured values). Since
the liquid-gas surface tension $\gamma _{LG}$ of the used liquids are known 
\cite{27, 34, 35}, and since the solid-gas surface tension $\gamma _{SG}$ of
chitinous material can be calculated (see further), it is possible to
calculate the contact angle $\theta _0 $ for a given liquid on a flat chitinous
surface, using the Young equation \cite{26}: 
\begin{equation}
\gamma _{SG}=\gamma _{SL}+\gamma _{LG}\cos \theta _0.  \label{young}
\end{equation}
Knowing that solid/liquid surface tension $\gamma _{SL}$ can be approximated
by \cite{26}: 
\begin{equation}
\gamma _{SL}\approx \gamma _{SG}+\gamma _{LG}-2\sqrt{\gamma _{SG}\gamma _{LG}%
},  \label{appro}
\end{equation}
we then get 
\begin{equation}
\cos \theta _0=2\sqrt{\frac{\gamma _{SG}}{\gamma _{LG}}}-1  \label{young2}
\end{equation}
from Eqs (3) and (4). Since $\gamma _{LG}$ of water is known, as well as the
contact angle of a water droplet on flat chitinous surface ($\theta
_0=105^{\circ }$) \cite{21}, we deduce from (3) $\gamma _{SG}\approx
9.9\times 10^{-3}$ N$\cdot $m$^{-1}$. In Fig. 10, gold and red lines denote
the linear asymptotic trends in data when $\cos \theta _{eff}<0$, i.e. when
the wing is not wet. By contrast, the blue line denotes the behavior of the
wet wing ($\cos \theta _{eff}>0$), i.e. the hydrophilic regime. The
horizontal green line corresponds to the process of hemiwicking \cite{30,31}%
. According to these results, we observe a transition from a
superhydrophobic regime to a hydrophilic regime.

\begin{figure}[t]
\centerline{\ \includegraphics[width=8.5 cm]{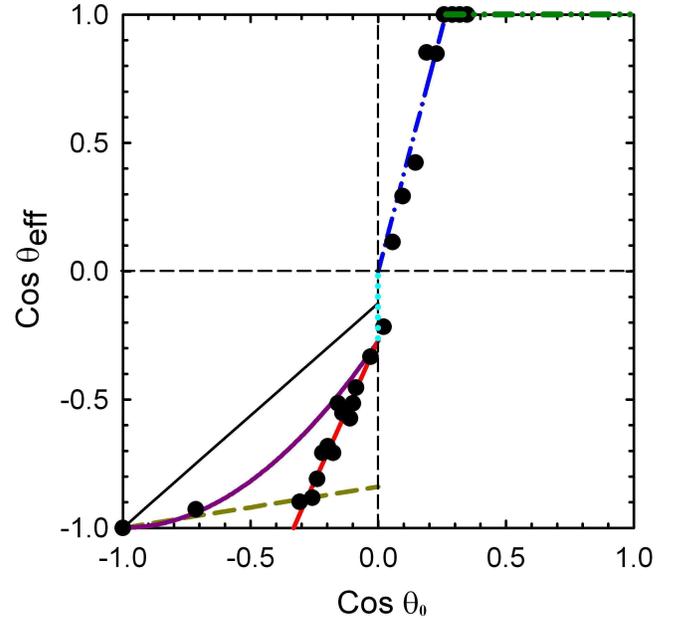}}
\caption{(Color online). Effective contact angle $\theta_{eff}$
(nanostructured surface) as a function of the contact angle $\theta_{0}$ on a
flat surface. The black dots are the experimental results. The blue
(dash-dotted) straight line is plotted from Eq. (9), the purple (solid)
curve is from Eq. (7). Gold (dashed) and red (solid) straight lines show
trends by contrast to the ideal model. Light blue (dotted) vertical straight
line corresponds to the hydrophobic/wetting transition. The green
(dash-dot-dotted) straight line corresponds to the process of hemiwicking. Black solid line is plotted from Eq. (8).}
\label{fig10}
\end{figure}

\subsection{Theoretical analysis}

Models of wetting processes are used hereafter in order to predict the
contact angle of a water droplet on the wing. Specifically, they are used to
describe the superhydrophobic, hydrophobic or hydrophilic behavior of the
wing. Two theoretical models are considered : the Cassie-Baxter model for
the hydrophobicity regime and Wenzel model for the hydrophilic regime.

The Cassie-Baxter (CB) model \cite{28} allows to describe the wetting
process on superhydrophobic and hydrophobic surfaces according to their
roughness (i.e. nipple array). For a surface made of two different materials
with different coverage areas, the contact angle $\theta_{CB}$ of the
heterogeneous surface is given by:

\begin{equation}
\cos {\theta _{CB}}=f_1\cos {\theta _1}+f_2\cos {\theta _2}.  \label{CB1}
\end{equation}
where $f_1$ and $f_2$ are the fractional areas of each material ($f_1$ + $%
f_2 $ $\geq$ 1), and $\theta _1$ and $\theta _2$ are the contact angles,
respectively.

Using the nipple geometry previously described, we consider the wetting
model shown on Fig. 9(a). The contact angle in the present case can be
written as \cite{28a}:

\begin{equation}
\cos {\theta _{CB}}=n\pi r^2(1+\cos {\theta }_0)^2-1,  \label{CB2}
\end{equation}
where $n$ is the number of hemispheres per unit area for a hexagonal array, $%
r$ is the radius of the hemisphere, and $\theta _0$ is the contact angle of
a droplet deposited on flat chitin. This well-known equation, can be simply
derived from Eq. (6) by assuming that $\theta _1=\theta $ and $\theta _2=180$%
$^{\circ }$. Indeed, $\theta _2$ is the contact angle between a droplet and
air, which is assumed to be $180$$^{\circ }$. Fractional areas can be easily
derived from trivial geometrical considerations. On Fig. 10, Eq. (7) is
plotted as the purple curve with ${\theta _{CB}}=\theta _{eff}$, using
geometrical parameters mentioned in section II and measured from SEM
pictures. Gold and red lines show wetting trends experimentally observed by
contrast to the theoretical model \cite{28b}. Though the theoretical curve
and the experimental data do not perfectly match, the trends are the same.
Differences are likely due to the fact that the water/air interface (see
Fig. 9(a)) is not truly a planar one, modifying then the conditions for
which the Gibbs energy is minimized \cite{26}. Nevertheless, more accurate
modeling is a complex task far beyond the scope of the present work \cite
{28c}.

We note that, using Eq. (7), we find a theoretical contact angle of $%
151^{\circ }$ for water by contrast with the measured value of $146^{\circ }$%
, i.e there is less than 4\% difference between theory and experiment.

In order to clearly demonstrate the essential role of the hemispherical top, we investigated
the wetting properties of the conical base structure without the top
hemisphere (Fig. 9(b)). In this case, the Cassie-Baxter contact angle is simply given by: 
\begin{equation}
\cos {\theta _{CB}}=f\cos {\theta _0}+\left( f-1\right) ,  \label{CB1b}
\end{equation}
with $f=\pi r^2/S$ and $S=a_0^2\sqrt{3}/2$. The linear dependence, which is plotted in Fig.
10 as a black solid line, leads to much less satisfactory agreement with data, as compared with the quadratic dependence (purple curve) pertaining to the 
hemispherical top, i.e. Eq. (7). Indeed, for water ($\theta _0$ = $105^{\circ }$), a theoretical contact angle of $\theta _{CB}$ = $111^{\circ }$ is predicted by 
Eq. (8), whereas a value of $151^{\circ }$ was determined previously by Eq. (7). Therefore, only the model including the hemispherical top is able to account for 
the superhydrophobicity of the wing.

The hydrophilic behavior of a corrugated surface, on the other hand, can be
described thanks to the Wenzel (W) model \cite{29}. In this case, the
contact angle is given by:

\begin{equation}
\cos {\theta _W}=\text{r}_{\text{o}}\cos {\theta }_0,  \label{W1}
\end{equation}
where r$_{\text{o}}$ is the roughness factor which is in our case given by:

\begin{equation}
\text{r}_{\text{o}}=\frac{2\pi r^2+\pi G(R+r)+a_0^2\sqrt{3}/2-\pi R^2}S,
\label{rou}
\end{equation}
with $S=a_0^2\sqrt{3}/2$, $r$ the radius of the hemisphere, $R$ the radius
of the base of the cone, $G$ the length of the cone ($\approx 166$ nm) and $a_0$ the spatial
period. Equation (8) is a linear relation between the effective contact
angle and the contact angle on a flat surface made of the same material. On
Fig. 10, Eq. (9) is plotted as blue curve with ${\theta _W}=\theta _{eff}$,
using geometrical parameters mentioned in section II and measured from SEM
pictures. Here, the model matches very well with the experimental data.

We emphasize the fact that the wetting models underline the role of the top
hemisphere in the wetting phenomena, and especially in the hydrophobic
behavior of the cicada wings. In the wetting regime, the whole structure
geometry must be considered to explain the wetting.

\section{Summary and conclusion}

While it is known that cicada wings possess both hydrophobic \cite{22,23,24}
and anti-reflective properties \cite{1,2}, in the present work we underlined
the two-in-one character of the geometrical features leading both to the
optical and wetting properties of the cicada. From scanning electron
microscopy, it was shown that the cicada wing consisted of a thick chitinous
layer whose each side was covered by a nanostructured device in the form of
an hexagonal nipple array. Each nipple was modeled by a truncated cone
topped by an hemisphere. We performed an original wetting experiment with
various chemical species as well as comprehensive optical measurements
providing the complex refractive index of the chitinous wing. From optical
simulations and using an analytical wetting model, we demonstrated the
character of the two functional levels of the nipple array. We concluded
that the hemisphere on the top of the nipple was useful for the
superhydrophobic character of the cicada wing. This half-sphere played no
optical role. By contrast, the truncated cone played no significant role in
the hydrophobic behavior while it was found to be fundamental for ensuring
transparency of the cicada wing (anti-reflection). Therefore, the nipple
array nanostructure in the grey cicada can be considered as a two-in-one
device: a hemisphere top which allows a strong water repellency and a
truncated cone which ensures anti-reflective properties.

\begin{acknowledgments}
The authors thank Corry Charlier, Jean-Fran\c cois Colomer, Quentin Spillier and Micha\"{e}l Lobet for their technical support and advice. L.D. is supported by the Belgian Fund for Industrial and Agricultural Research (FRIA). M.S. is supported by 
the Cleanoptic project (development of super-hydrophobic anti-reflective coatings for solar glass panels/convention No.1117317) of the Greenomat program of the 
Wallonia Region (Belgium).
\end{acknowledgments}

\end{document}